\documentclass[a4paper,longbibliography,english,reprint,aps,superscriptaddress,nofootinbib]{revtex4-2}

\usepackage{babel,calc,amsmath,amsthm,amssymb,bm,graphicx,siunitx}
\graphicspath{{images/}{images/SM/}}

\usepackage[T1]{fontenc}
\usepackage{mathdots}
\usepackage{geometry}
\usepackage[unicode=true]{hyperref}
\usepackage{booktabs}
\usepackage{mathtools}
\usepackage{multirow}
\usepackage{lipsum}
\usepackage[normalem]{ulem}
\usepackage[dvipsnames]{xcolor}
\usepackage{placeins}

\usepackage[ruled]{algorithm2e}
\SetKwInput{KwData}{Input}
\SetKwInput{KwResult}{Output}
\SetKwComment{Comment}{~\#~}{}

\hypersetup{
     colorlinks=true,               
     linkcolor=Blue,            
     citecolor=Blue,             
     urlcolor=Blue,          
 }

\newtheorem*{corollary*}{Corollary}

\newtheorem*{proposition*}{Proposition}

\newtheorem*{theorem*}{Theorem}

\theoremstyle{definition}

\newtheorem*{definition*}{Definition}

\theoremstyle{remark}

\newtheorem*{lemma*}{Lemma}

\newtheorem*{remark*}{Remark}

\newtheorem*{example*}{Example}

\newif\ifdebug

\debugtrue

\ifdebug

\newcommand\delete{\bgroup\markoverwith{\textcolor{Maroon}{\rule[0.5ex]{2pt}{0.4pt}}}\ULon}

\else

\newcommand{\note}[1]{\ignorespaces}
\newcommand{\delete}[1]{\ignorespaces}

\fi

\begin{document}
\renewcommand{\figurename}{Fig.}

\newcommand{\extfig}{Extended Data Figure}
\newcommand{\methodsname}{Methods}
\newcommand{\smname}{Supplementary Material}

\renewcommand{\sectionautorefname}{Sec.}
\renewcommand{\tableautorefname}{Table}
\renewcommand{\equationautorefname}{Eq.}


\title{{Wave--particle transition and quantum Zeno effect in which-way experiments with a superconducting quantum processor}}

\author{Shiyu~Wang}
\email{shiyu.wang@riken.jp}
\affiliation{RIKEN Center for Quantum Computing (RQC), Wako, Saitama 351-0198, Japan}

\author{Zhiguang~Yan}
\email{zhiguang.yan@riken.jp}
\affiliation{RIKEN Center for Quantum Computing (RQC), Wako, Saitama 351-0198, Japan}

\author{Clemens~Gneiting}
\author{Rui~Li}
\affiliation{RIKEN Center for Quantum Computing (RQC), Wako, Saitama 351-0198, Japan}

\author{Franco~Nori}
\affiliation{RIKEN Center for Quantum Computing (RQC), Wako, Saitama 351-0198, Japan}
\affiliation{Department of Physics, University of Michigan, Ann Arbor, Michigan 48109-1040, USA}

\author{Yasunobu~Nakamura}
\affiliation{RIKEN Center for Quantum Computing (RQC), Wako, Saitama 351-0198, Japan}
\affiliation{Department of Applied Physics, Graduate School of Engineering, The University of Tokyo, Bunkyo-ku, Tokyo 113-8656, Japan}


\begin{abstract}
    Wave--particle duality demonstrates the peculiar nature of quantum mechanics. In which-way experiments, depending on the measurement scheme, a particle exhibits either wave-like or particle-like properties, as summarized by Bohr’s principle of complementarity. In this work, we implement Mach–Zehnder (MZ) interferometry on a two-dimensional (2D) superconducting quantum processor. With precise control of the which-way measurement strength, we demonstrate the transition of a photon from wave-like to particle-like behavior. Furthermore, by performing quantum state tomography on two qubits located in the two paths, we demonstrate that which-way measurements break the entanglement and coherence between the two paths and cause information leakage from the quantum system to the environment. To capture this behavior quantitatively, we derive complementarity relations between the entropy and the fringe visibility. By applying a continuous which-way measurement during the evolution, we also observe the quantum Zeno effect that partially obstructs the interferometer path, giving rise to nonmonotonic behavior of purity and von Neumann entropy. Our experiments provide a detailed characterization of the full interferometer dynamics, reveal the relation between wave--particle duality and quantum information, and demonstrate the potential of superconducting quantum processors for testing quantum foundations under high precision and controllability.
\end{abstract}

\maketitle


Bohr’s complementarity principle~\cite{bohr1928quantenpostulat,englert1995complementarity,schleich2016wave} represents a fundamental principle of quantum mechanics that reveals its profound essence. It asserts that a quantum object can behave like a wave or a particle depending on the choice of measurement, while these two mutually exclusive behaviors cannot be observed in the same experiment. Which-way experiments are a paradigmatic class of complementarity demonstrations that includes double-slit and MZ interferometry~\cite{scully1989quantum, sanders1989complementarity, scully1991quantum,scully1997quantum,agarwal2012quantum}. These have been realized in various platforms, such as photonic~\cite{rarity1990two,kim2000delayed,mir2007double}, electronic~\cite{ji2003electronic,johnson2022inelastic} and atomic~\cite{carnal1991young,durr1998origin} systems. In a two-way interferometer, if no which-way information is acquired, the wave-like behavior is demonstrated by a high-visibility interference pattern. On the other hand, if which-way information is collected, the interference fringes disappear, indicating particle‐like behavior. In the intermediate regime, where incomplete which-way information is retrieved, the interference fringes still persist but with a reduced visibility, giving rise to a complementarity relation~\cite{englert1996fringe,durr1998fringe,chen2022experimental}: 
\begin{align} 
D^2 + V^2 \le 1,
\end{align} 
where $D$ is the distinguishability of the ways through the interferometer and $V$ is the visibility of the interference fringes.
Moreover, the quantum wave--particle superposition state can be realized in a single experiment~\cite{tang2012realization,auccaise2012experimental,qin2019proposal,wang2019quantum}. Further theoretical work and experiments have extended complementarity relations to the coherence~\cite{mandel1991coherence,bagan2016relations,gao2018experimental} and entropy~\cite{coles2014equivalence,bagan2020wave} of quantum systems, unveiling the fundamental connection between complementarity and quantum information.

Owing to their scalability and programmability, superconducting-qubit systems provide an excellent platform for quantum simulations~\cite{houck2012chip,georgescu2014quantum,google2020hartree,kim2023evidence}. In prior superconducting qubit experiments, simple proof-of-principle demonstrations of MZ interferometers have been realized~\cite{oliver2005mach,gong2021quantum,zhang2022synthesizing}. In optical systems, detecting the position of a photon typically requires absorbing the photon~\cite{komiyama2000single,hadfield2009single,moreau2017demonstrating}. However, in superconducting-qubit systems, by performing quantum nondemolition (QND)~\cite{wallraff2004strong,kono2018quantum,gusenkova2021quantum} measurements on the qubits, we can determine the photon position without destroying it and also record its entire propagation path. Furthermore, by varying the strength of the dispersive measurement, we can readily tune the which-way measurement from the weak regime to a projective regime~\cite{hatridge2013quantum,murch2013observing}. This enables us to investigate the impact of the which-way measurement strength on the interference fringes and the quantum information of the system.

Here, we demonstrate MZ interferometry with a 4-qubit and a 12-qubit system. We first generate a microwave photon by exciting a single qubit with a $\pi$ pulse. The photon propagates and interferes along the two arms of an MZ interferometer composed of the qubits. A qubit in one of the arms acts as the which-way detector. By gradually increasing the measurement strength of the detector, we observe a progressive reduction in interference-fringe visibility, demonstrating the transition of the microwave photon from wave-like to particle-like behavior. From the quantum state tomography of the two-qubit system formed by the qubits in the two arms, we observe that, as the measurement strength increases, the von Neumann entropy increases while the concurrence and purity decrease. We derive inequalities between the purity, von Neumann entropy and fringe visibility, demonstrating the tight intertwining between complementarity and quantum information. Under continuous measurement, we further observe the quantum Zeno effect accompanied by nonmonotonic behavior of the purity and von Neumann entropy, providing insight on time-of-arrival measurements.

\begin{figure*}[t]
    \centering
    \includegraphics[width=1.0\textwidth]{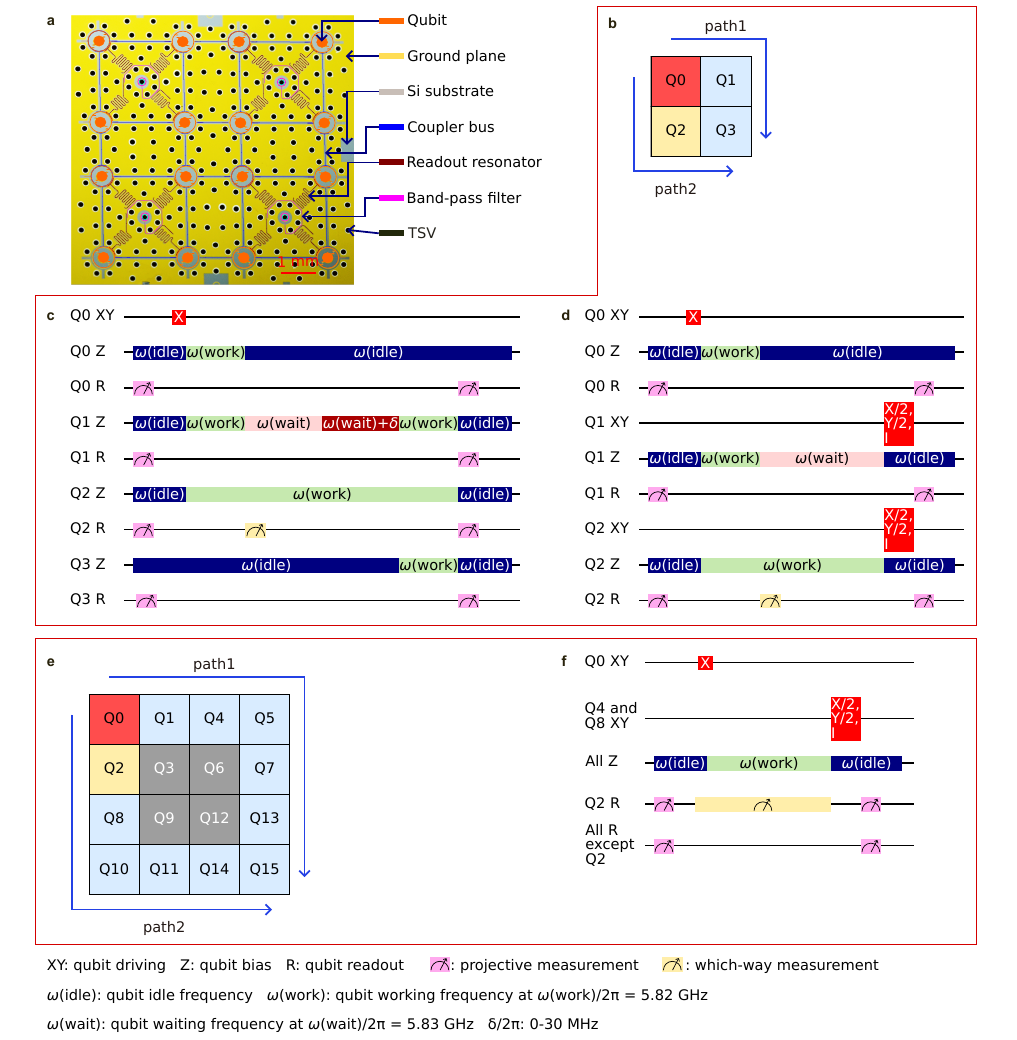}
    \caption{\textbf{Experimental setup and quantum circuits.}
		\textbf{a,} ~False-colored optical image of the 16-qubit superconducting chip. The 16-qubit chip has a $4\times4$ qubit array composed of four units; each unit contains four frequency-tunable transmons (orange). Each qubit is coupled to an individual $\lambda/4$ readout resonator (brown), and the four readout resonators within the same unit are coupled to a common Purcell filter (pink). Neighboring qubits are coupled via a coupler bus (blue lines). The chip contains many through-silicon vias (TSV, black dots). Qubits are controlled and read out through pogo pins connected to coaxial cables from the backside of the chip.
		\textbf{b,} Qubit configuration of the 4-qubit Mach–Zehnder (MZ) interferometer.
        \textbf{c,} Quantum-circuit diagram for the 4-qubit MZ interferometer.
		\textbf{d,} Quantum circuit for the 2-qubit tomography experiment for the 4-qubit MZ interferometer.
        \textbf{e,} Qubit configuration of the 12-qubit MZ interferometer. 
		\textbf{f,}~Quantum circuit for the 12-qubit MZ interferometer.
        In both \textbf{b} and \textbf{e}, a single photon is generated by exciting Q0. Then the photon propagates along path~1 and path~2, respectively. The which-way measurement is applied to Q2.
        }

    \label{fig_1}
\end{figure*}

We realize the 4-qubit and 12-qubit MZ interferometers with a quantum processor comprised of 16 frequency-tunable transmon qubits~\cite{yan2026characterizing}. As shown in Fig.~\ref{fig_1}a, our processor features a scalable three-dimensional vertical wiring architecture~\cite{yan2026characterizing,spring2025fast}, with control and readout signals delivered via pogo pins from the backside. The quantum chip contains four parallel readout units, each addressing four transmon qubits. Each qubit has an individual readout resonator, and the four resonators within the same unit are coupled to a common Purcell filter. Neighboring qubits are capacitively coupled via a coupler bus. The Hamiltonian of our system can be described by a Bose-Hubbard model:
\begin{eqnarray}
\hat H &=& \sum_i \left[\hbar\omega_i \hat a_i^{\dagger}\hat a_i
+ \frac{\hbar U_i}{2}\hat n_i(\hat n_i-1)\right] \nonumber\\
&& + \sum_{\langle i,j\rangle}\hbar J_{ij}
\left(\hat a_i^{\dagger}\hat a_j+\hat a_i\hat a_j^{\dagger}\right),
\label{H}
\end{eqnarray}	
where $\omega_i$ is the qubit frequency, $a_i^{\dagger}$ ($a_i$) the bosonic creation (annihilation) operator, $U_i$ the anharmonicity of the qubit, $\hat n_i = \hat a_i^{\dagger}\hat a_i$ the number operator, and $J_{ij}$ the photon hopping rate between nearest-neighbor qubits Q$i$ and Q$j$. For the transmon qubits in our system, $J_{ij} \ll U_i$, the system Hamiltonian can be mapped to a hard-core-boson (HCB) Hamiltonian~\cite{yanay2020two,karamlou2024probing}: 
\begin{align} \label{H_HCB}
	\hat H_\mathrm{HCB} = \sum _{i}\frac{\hbar\omega_i}{2}\hat\sigma_i^z + \sum _{\langle i,j\rangle}\hbar J_{ij}(\hat\sigma_i^+\hat\sigma_j^-+\hat\sigma_i^-\hat\sigma_j^+),
\end{align}	
where $\hat\sigma_i^z$ is the Pauli Z operator and $\hat\sigma_i^+$ ($\hat\sigma_i^-$) is the qubit raising (lowering) operator. It is equivalent to the spin-$1/2$ antiferromagnetic XY model.

\begin{figure*}[t]
    \centering
    \includegraphics[width=1.0\textwidth]{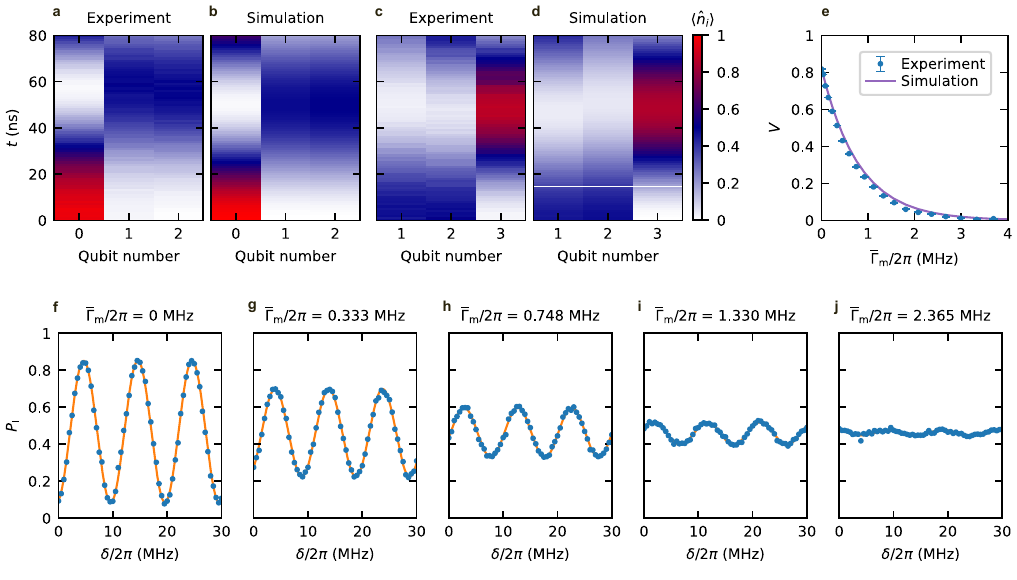}
    \caption{\textbf{Evolution and interference pattern of the 4-qubit MZ interferometer.} \textbf{a,} Evolution of the population $\langle \hat n_i \rangle$ in Q$i$~($i=0,1,2$) from  $t=0$~ns to $t=80$~ns. The photon initially generated at Q0 propagates to Q1 and Q2. For the subsequent which-way measurement experiments, we decouple the qubits when the population of Q0 reaches its minimum.  \textbf{b,} Numerical simulation of the population evolution under the same
conditions as in \textbf{a}. \textbf{c,} Evolution of $\langle \hat n_i \rangle$~($i=1,2,3$) from  $t=0$~ns to $t=80$~ns. The photon that propagates to Q1 and Q2 interferes at Q3.  \textbf{d,} Numerical simulation of the population evolution under the same
conditions as in \textbf{c}.
\textbf{e,} Relation between the interference visibility $V$ of the 4-qubit MZ interferometer and $\overline{\Gamma}_\mathrm{m}/2\pi$.
		\textbf{f--j,} Interference pattern produced by the 4-qubit MZ interferometer under the average which-way measurement-induced dephasing rates $\overline{\Gamma}_\mathrm{m}/2\pi$ = 0, 0.333, 0.748, 1.330, and 2.365 MHz, respectively. $\delta$ is the detuning of Q1 from 5.83 GHz during the $Z$ pulse with the length of $t_\mathrm{p} =100$~ns, $P_\mathrm{I}$ is the population of Q3 when the photon interferes at it. We fit the experimental data (blue dots) with a sinusoidal curve (orange curve). 
	}
    \label{fig_2}
\end{figure*}

We first implement an MZ interferometer in a system comprising four qubits, Q0, Q1, Q2, and Q3, as shown in Fig.~\ref{fig_1}b. Figure~\ref{fig_1}c shows the quantum circuit used in the experiment. We first bias all qubits to their idle frequencies. Initial measurements of all qubits are performed to post-select the initial state of each qubit to be state $\lvert 0\rangle$. We generate a photon in the system by exciting Q0. We then tune Q0, Q1, and Q2 to the same working frequency of \SI{5.82}{\giga\hertz} by applying $Z$ pulses, allowing the photon to propagate from Q0 to Q1 and Q2. Figure~\ref{fig_2}a shows the evolution of the population in the three-qubit system, and Fig.~\ref{fig_2}b is the simulation result. The simulations in this paper are done with QuTiP~\cite{johansson2012qutip,LAMBERT20261}. Q0 acts as the beam splitter of the MZ interferometer. After the beam splitter the photon propagates along Q1 (path 1) and Q2 (path 2). When the population of Q0 reaches its minimum at \SI{57}{\nano\second}, we bias Q0 to its idle frequency, keep Q2 at the working frequency and bias Q1 to the waiting frequency of \SI{5.83}{\giga\hertz}. By detuning the frequencies of the nearest-neighbor qubits, we decouple the qubits to freeze the population transfer. Then we apply a 100\nobreakdash-\si{\nano\second} dispersive measurement pulse of amplitude $A$ to Q2's readout resonator. This constitutes a which-way measurement on path 2. We then wait 100 ns for the photons in the readout resonator to decay and apply a $Z$ pulse to bias Q1 to $\omega_\mathrm{Q1}/2\pi = \SI{5.83}{\giga\hertz} + \delta/2\pi$ for $t_\mathrm{p}=\SI{100}{\nano\second}$, which causes the photon to accumulate an extra phase $\phi = -t_\mathrm{p}\delta$ in path 1. We vary $\phi$ by adjusting $\delta/2\pi$ from 0 to \SI{30}{\mega\hertz} to produce interference fringes. Next, we bias Q1, Q2, and Q3 to \SI{5.82}{\giga\hertz} with $Z$ pulses to let them evolve, where Q3 serves as the second beam splitter of the MZ interferometer. Figure~\ref{fig_2}c shows the evolution of the population in the system of Q1, Q2, and Q3 at this stage, and Fig.~\ref{fig_2}d is the simulation result. Finally, we measure Q3 when its population $P_\mathrm{I}$ reaches the maximum after 49-ns evolution to obtain the interference fringes shown in Fig.~\ref{fig_2}f.

By first preparing Q2 in the state $\bigl(\lvert 0\rangle-i\lvert 1\rangle\bigr)/\sqrt2$ and performing single-qubit state tomography on Q2 after applying the measurement, we calibrate the average measurement-induced dephasing rate $\overline{\Gamma}_\mathrm{m}$ of Q2 in the 200-ns period~\cite{supplementary_mypaper}. When we increase $\overline{\Gamma}_\mathrm{m}/2\pi$ from 0 to 3.695~MHz, path 2 transitions from no which-way measurement through a weak which-way measurement to a projective which-way measurement. In Figs.~\ref{fig_2}f--j, we observe that, as the measurement strength increases, the interference pattern becomes weaker until it finally disappears. We define the visibility $V$ as
\begin{align} \label{visibility}
	V = \frac{P_\mathrm{Imax}-P_\mathrm{Imin}}{P_\mathrm{Imax}+P_\mathrm{Imin}}.
\end{align}	
Figure~\ref{fig_2}e shows that the visibility decreases with $\overline{\Gamma}_\mathrm{m}$ and finally approaches 0. The experimental result agrees well with the simulation, indicating that the disappearance of the interference fringes in which-way experiments can be well explained by the measurement-induced dephasing. The measurement entangles the photon in the system with the measurement apparatus, causing dephasing of the photon and leakage of quantum information into the measurement apparatus. 

\begin{figure*}[t]
    \centering
    \includegraphics[width=1.0\textwidth]{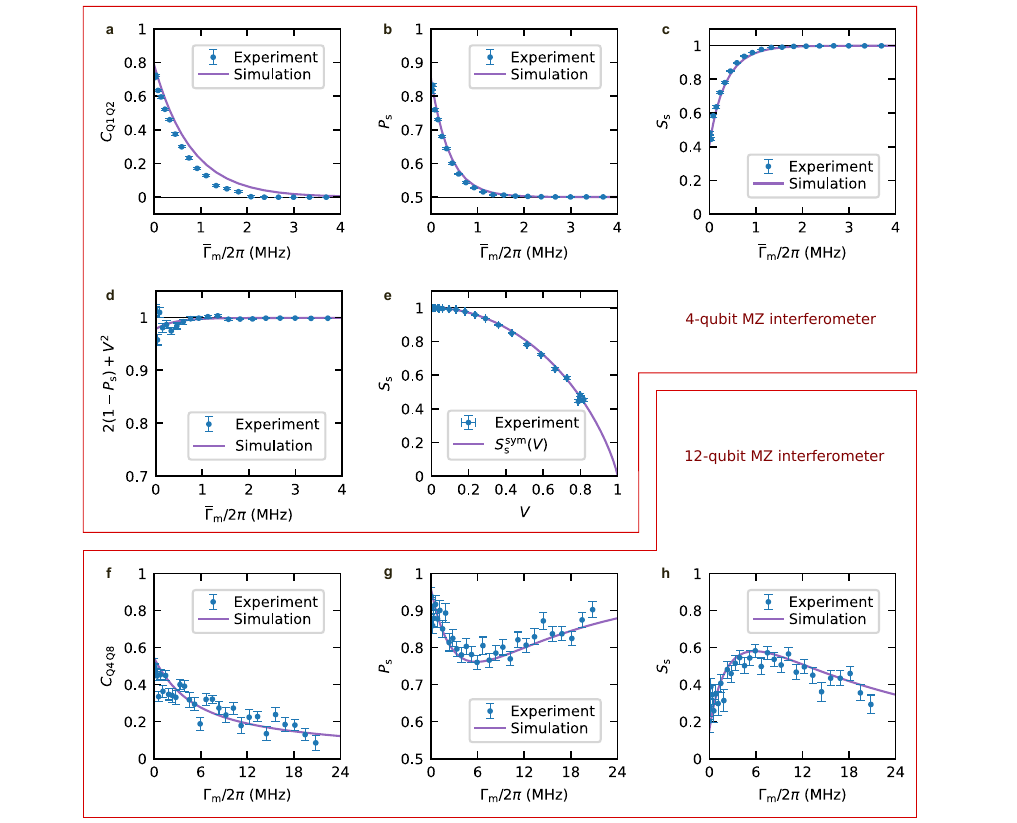}
    \caption{\textbf{Relations between quantum information, which-way measurement-induced dephasing rate, and visibility of the MZ interferometers.} \textbf{a}--\textbf{e} refer to the 4-qubit MZ interferometer, while \textbf{f}--\textbf{h} correspond to the 12-qubit MZ interferometer.
		\textbf{a,}~Concurrence $C_\mathrm{Q1Q2}$ of Q1 and Q2 vs.\ the average which-way measurement-induced dephasing rate $\overline{\Gamma}_\mathrm{m}$. 
		\textbf{b,}~Purity $P_\mathrm{s}$ in the single-excitation subspace vs.\ $\overline{\Gamma}_\mathrm{m}$. 
		\textbf{c,}~Von Neumann entropy $S_\mathrm{s}$ in the single-excitation subspace vs.\ $\overline{\Gamma}_\mathrm{m}$. 
		\textbf{d,}~Complementarity relation between linear entropy $1-P_{\mathrm{s}}$ and visibility $V$.        
		\textbf{e,}~$S_\mathrm{s}$ vs.\ $V$. The purple curve represents the theoretical $S_\mathrm{s}^\mathrm{sym}(V)$ given by Eq.~\eqref{inequality_S}.     
            \textbf{f,}~Concurrence $C_\mathrm{Q4Q8}$ of Q4 and Q8 vs.\ the which-way measurement-induced dephasing rate $\Gamma_\mathrm{m}$.
            \textbf{g,}~$P_\mathrm{s}$ vs.\ $\Gamma_\mathrm{m}$.
            \textbf{h,}~$S_\mathrm{s}$ vs.\ $\Gamma_\mathrm{m}$.            
	}
    \label{fig_4}
\end{figure*}

To further illuminate the origin of the which-way measurement’s effect on the interference pattern, we slightly modify the quantum circuit in Fig.~\ref{fig_1}c to obtain Fig.~\ref{fig_1}d: after performing the which-way measurement on Q2, we do not accumulate the phase $\phi$ nor allow the joint evolution of Q1, Q2, and Q3; instead we wait 200 ns and then perform two-qubit state tomography on Q1 and Q2. With the reconstructed two-qubit density matrix $\rho_\mathrm{Q1Q2}$, we compute the two-qubit concurrence~\cite{wootters1998entanglement} 
\begin{align} \label{concurrence}
C_\mathrm{Q1Q2} = \max\bigl(0,\;\lambda_1-\lambda_2-\lambda_3-\lambda_4\bigr),
\end{align}
in which $\lambda_i$s are the square roots of the eigenvalues of the non-Hermitian matrix $R=\rho\,\tilde{\rho}$ in decreasing order, where $\tilde{\rho}=(\sigma_y\otimes\sigma_y)\,\rho^*\,(\sigma_y\otimes\sigma_y)$, and $\rho^*$ is the complex conjugate of $\rho$.
As shown in Fig.~\ref{fig_4}a, when increasing the which-way measurement strength, the two-qubit concurrence falls from about 0.717 to nearly 0, indicating progressive destruction of the entanglement between the two qubits in the two paths. 

We then calculate the two-qubit density matrix restricted to the single-excitation subspace by
\begin{align} \label{rho_sub}
\rho_\mathrm{s} = \frac{\hat P\,\rho_\mathrm{Q1Q2}\,\hat P}{\mathrm{Tr}\bigl(\hat P\,\rho_\mathrm{Q1Q2}\bigr)},
\end{align}
where $\hat P = \lvert 10\rangle\langle 10\rvert + \lvert 01\rangle\langle 01\rvert$ is the single-excitation projector. Figure~\ref{fig_4}b shows that the which-way measurement degrades the purity of the system, $P_\mathrm{s} = \operatorname{Tr}(\rho_\mathrm{s}^2)$, driving the state toward a mixed state. Figure~\ref{fig_4}c shows  that the which-way measurement increases the entropy of entanglement between the system and the environment, which is equal to the von Neumann entropy of the system, $S_\mathrm{s} = -\mathrm{Tr}(\rho_\mathrm{s}\log\rho_\mathrm{s})$, thereby leaking quantum information into the measurement apparatus.

We then derive an inequality between the purity in the single-excitation subspace and the visibility~\cite{supplementary_mypaper},
\begin{align} \label{inequality_P}
2(1-P_\mathrm{s}) + V^{2} \;\le\; 1,
\end{align}
and an inequality between the von Neumann entropy in the single-excitation subspace and the visibility,
\begin{eqnarray}
S_\mathrm{s} &\le& S_\mathrm{s}^\mathrm{sym}(V) \nonumber\\
&=& -\biggl(\frac{1+V}{2}\biggr)\log_{2}\biggl(\frac{1+V}{2}\biggr) \nonumber\\
&& - \biggl(\frac{1-V}{2}\biggr)\log_{2}\biggl(\frac{1-V}{2}\biggr),
\label{inequality_S}
\end{eqnarray}
where $S_\mathrm{s}^\mathrm{sym}(V)$ is the von Neumann entropy when the two paths are symmetric. The equality in both inequalities holds when the two paths of the interferometer are symmetric, a condition that is approximately satisfied in our 4-qubit MZ interferometer. Note that $(1-P_\mathrm{s})$ is the linear entropy and gives a direct measure of the mixedness of the quantum state~\cite{zurek1993coherent,isar1999purity}. Equation~\eqref{inequality_P} demonstrates the complementarity relation between the linear entropy and the visibility, as shown in Fig.~\ref{fig_4}d. Equation~\eqref{inequality_S} relates the von Neumann entropy to the visibility, and Fig.~\ref{fig_4}e shows that $S_\mathrm{s}^\mathrm{sym}(V)$ provides a good fit to the experimental data. These two inequalities quantitatively characterize the relationship between information leakage in the quantum system and wave--particle duality.

\begin{figure*}[t]
    \centering
    \includegraphics[width=1.0\textwidth]{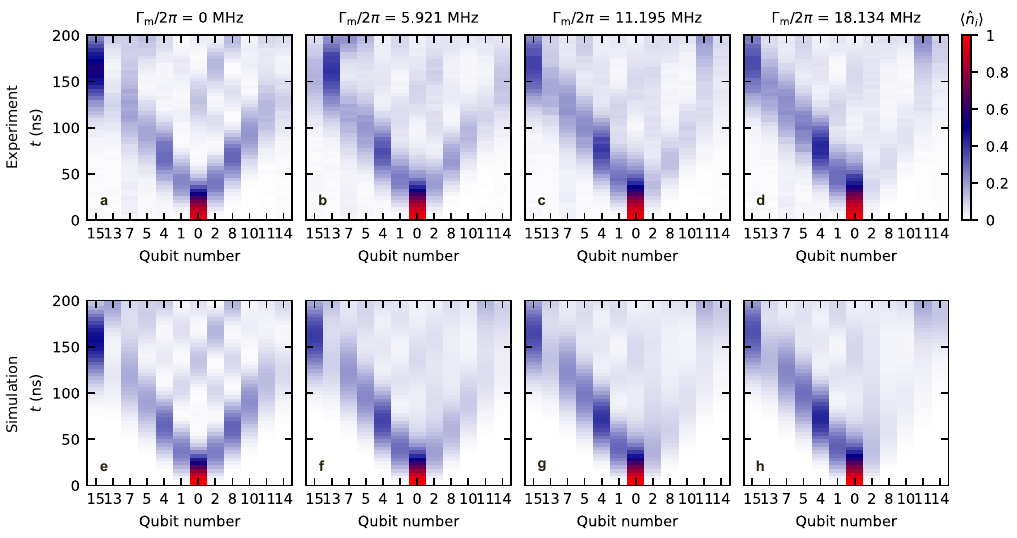}
    \caption{\textbf{Quantum Zeno effect in the 12-qubit MZ interferometer.}
		\textbf{a}--\textbf{d,} Evolution of Q$i$'s population $\langle \hat n_i \rangle$ from  $t=0$~ns to $t=200$~ns under the which-way measurement-induced dephasing rate $\Gamma_\mathrm{m}/2\pi$ = 0, 5.921, 11.195, and 18.134~MHz, respectively. Path~2 of the MZ interferometer is blocked by the quantum Zeno effect as the which-way measurement strength applied to Q2 increases.
		\textbf{e}--\textbf{h,} Numerical simulations of the population evolution under the same conditions as in \textbf{a}--\textbf{d}.  
	}
    \label{fig_3}
\end{figure*}

We next extend the MZ interferometer to a larger system and a continuous measurement scheme. We construct an MZ interferometer by the 12 qubits shown in Fig.~\ref{fig_1}e, and the quantum circuit is depicted in Fig.~\ref{fig_1}f. First, we bias all qubits to their idle frequencies and generate a photon by exciting Q0. We then tune the 12 qubits that form the interferometer to the working frequency of $\SI{5.82}{\giga\hertz}$ for evolution. After the beam splitter Q0, the photon propagates along two paths: path 1 formed by Q1, Q4, Q5, Q7, and Q13, and path 2 formed by Q2, Q8, Q10, Q11, and Q14. Figure~\ref{fig_3}a shows the evolution of the population in the 12-qubit system, and Fig.~\ref{fig_3}e shows the simulation. Finally, the photon interferes at the beam splitter Q15. In contrast to the 4-qubit MZ interferometer, where the which-way measurements are performed while the qubits are decoupled, the 12-qubit experiment uses a continuous measurement scheme. We start applying the which-way measurement pulse to Q2 before the initial state preparation so that the measurement strength is already stabilized when the 12-qubit joint evolution begins. We maintain this which-way measurement throughout the evolution and compensate the Stark shift induced by the which-way measurement pulse with the $Z$-bias pulse on Q2.

We observe the effect of the which-way measurement on the MZ interference by sweeping the strength of the which-way measurement applied to Q2. We calibrate the measurement-induced dephasing rate $\Gamma_\mathrm{m}$ by Stark-shift experiments.  Figures~\ref{fig_3}a--d show the system evolution over time for different $\Gamma_\mathrm{m}/2\pi$, from 0 to 18.134~MHz, and Figs.~\ref{fig_3}e--h show the respective numerical simulations. We can see that, when no measurement is applied, the photon propagates along both paths 1 and 2, and then interferes at the beam splitter Q15. As the measurement strength increases, the population in path 2 gradually decreases, and eventually the photon predominantly propagates to Q15 via path 1. This behavior arises because of the which-way measurement-induced quantum Zeno effect~\cite{gambetta2008quantum,matsuzaki2010quantum,dhar2015detection,patil2015measurement} that tends to keep Q2 in the initial state $\lvert 0\rangle$ and to reflect the propagating photon; the stronger the measurement, the stronger the induced reflection.
This observation is also consistent with approaches to determine the photon arrival time~\cite{dhar2015detection,aharonov1998measurement,muga2000arrival,dubey2021quantum}: in order to measure the precise time the photon arrives one must monitor the arrival region. If the continuous measurement is too weak, it fails to obtain accurate arrival-time information, whereas if the measurement is too strong, it reflects the photon and thus prevents detection. Only at intermediate measurement strengths one can both pass the photon and measure its arrival time with sufficient accuracy.

To analyze the quantum-information leakage, we perform two-qubit state tomography on Q4 and Q8 while varying the which-way measurement-induced dephasing rate $\Gamma_\mathrm{m}$ of Q2. We then extract the two-qubit concurrence $C_\mathrm{Q4Q8}$, the purity $P_\mathrm{s}$ and the von Neumann entropy $S_\mathrm{s}$ in the single-excitation subspace at the time when $C_\mathrm{Q4Q8}$ reaches its maximum in the absence of which-way measurement. $C_\mathrm{Q4Q8}$ falls with increasing $\Gamma_\mathrm{m}$ (Fig.~\ref{fig_4}f), demonstrating progressive destruction of entanglement between the two paths. In contrast to the 4-qubit MZ interferometer, $P_\mathrm{s}$ exhibits a nonmonotonic dependence on measurement strength (falling then rising) (Fig.~\ref{fig_4}g), and $S_\mathrm{s}$ shows the opposite nonmonotonic trend (rising then falling) (Fig.~\ref{fig_4}h). These observations imply that, once the measurement-induced dephasing rate greatly exceeds the Q0--Q2 hopping rate (3.56 MHz), the quantum Zeno effect dominates: photons are reflected and the information leaking into the measurement apparatus is reduced. When the purity reaches its minimum and the von Neumann entropy reaches its maximum, information leakage is maximal. This provides a possible method to identify the optimal measurement strength in time-of-arrival measurements based on the von Neumann entropy and the purity.

In summary, we realized two MZ interferometers---one comprising 4 qubits and the other 12 qubits---on a 2D 16-qubit superconducting processor. We implemented which-way measurements both for coupled and decoupled qubits and investigated how the resulting measurement-induced dephasing and the quantum Zeno effect affect the interference. By performing simultaneous readout of all qubits, we continuously monitored the full MZ interferometer dynamics and observed the measurement-induced wave-to-particle transition. We extracted the two-qubit concurrence, the purity and the von Neumann entropy in the single-excitation subspace as functions of the which-way measurement-induced dephasing rate, demonstrating that the which-way measurement destroys inter-path entanglement and leaks quantum information into the measurement apparatus. Under continuous measurement, when the quantum Zeno effect prevails, photon reflection suppresses information leakage. We further derived inequalities between the purity, the von Neumann entropy and the fringe visibility, establishing a connection between wave–particle duality and quantum information and enriching the understanding of Bohr's complementarity. This work highlights the potential of programmable superconducting platforms for realizing and improving interference and other fundamental experiments. In future work, the setup can be adapted to implement, for instance, delayed-choice experiments~\cite{wheeler1978past,hellmuth1987delayed,peruzzo2012quantum,tang2012realization,dong2020temporal}, Franson Bell-CHSH experiments~\cite{franson1989bell,aerts1999two,marquardt2007efficient,gneiting2008bell,cabello2009proposed}, and multi-photon interference~\cite{tillmann2015generalized,peropadre2016proposal,agne2017observation}  on superconducting quantum processors.


\let\oldaddcontentsline\addcontentsline
\renewcommand{\addcontentsline}[3]{}
\bibliographystyle{rev4-2mod}
\bibliography{main}
\let\addcontentsline\oldaddcontentsline


\vspace{8pt} \noindent \textsf{\textbf{Acknowledgements}} - S.W., Z.Y., R.L. and Y.N. are partially supported by Ministry of Education, Culture, Sports, Science and Technology (MEXT) Quantum Leap Flagship Program (Q-LEAP)
    (via Grant No. JPMXS0118068682) and the Japan Science and Technology Agency (JST) as part of Adopting Sustainable Partnerships for Innovative Research Ecosystem (ASPIRE) (Grant No.~JPMJAP2513). C.G. and F.N. are supported in part by:
      the Japan Science and Technology Agency (JST)
       [via the CREST Quantum Frontiers program Grant No. JPMJCR24I2,
      the Quantum Leap Flagship Program (Q-LEAP), and the Moonshot R\&D Grant Number JPMJMS2061],
      and the Office of Naval Research (ONR) Global (via Grant No. N62909-23-1-2074).
    
\vspace{5pt} \noindent \textsf{\textbf{Author contributions}} - S.W. and Z.Y. conceived the research. 
    F.N. and Y.N. supervised the project.
    S.W. and Z.Y. performed the experiments.
    S.W. and C.G. contributed to the underlying theory.
    S.W. performed the data analysis and numerical simulations.
    Z.Y. designed and fabricated the sample. 
    Z.Y. set up the measurement system.
    S.W., R.L., and Z.Y. developed the measurement code. 
	S.W. wrote the manuscript.
	All authors contributed to the discussion of the results and to revising the manuscript. 

 \vspace{5pt} \noindent \textsf{\textbf{Competing interests}} - The authors declare no competing interests.

\end{document}



\title{{Supplementary Material for ``Wave–particle transition and quantum Zeno effect in which-way experiments with a superconducting quantum processor''}}

\author{Shiyu~Wang}
\email{shiyu.wang@riken.jp}
\affiliation{RIKEN Center for Quantum Computing (RQC), Wako, Saitama 351-0198, Japan}

\author{Zhiguang~Yan}
\email{zhiguang.yan@riken.jp}
\affiliation{RIKEN Center for Quantum Computing (RQC), Wako, Saitama 351-0198, Japan}

\author{Clemens~Gneiting}
\author{Rui~Li}
\affiliation{RIKEN Center for Quantum Computing (RQC), Wako, Saitama 351-0198, Japan}

\author{Franco~Nori}
\affiliation{RIKEN Center for Quantum Computing (RQC), Wako, Saitama 351-0198, Japan}
\affiliation{Department of Physics, University of Michigan, Ann Arbor, Michigan 48109-1040, USA}

\author{Yasunobu~Nakamura}
\affiliation{RIKEN Center for Quantum Computing (RQC), Wako, Saitama 351-0198, Japan}
\affiliation{Department of Applied Physics, Graduate School of Engineering, The University of Tokyo, Bunkyo-ku, Tokyo 113-8656, Japan}

\maketitle

\renewcommand{\figurename}{Fig.}

\newcommand{\extfig}{Extended Data Figure}
\newcommand{\methodsname}{Methods}
\newcommand{\smname}{Supplementary Material}

\renewcommand{\sectionautorefname}{Sec.}
\renewcommand{\tableautorefname}{Table}
\renewcommand{\equationautorefname}{Eq.}


\appendix
\onecolumngrid

\setcounter{equation}{0}
\counterwithout{equation}{section}
\setcounter{figure}{0}
\renewcommand{\theequation}{S\arabic{equation}}
\renewcommand{\thefigure}{S\arabic{figure}}
\renewcommand{\theHequation}{S\arabic{equation}}
\renewcommand{\theHfigure}{S\arabic{figure}}
\renewcommand{\appendixname}{}
\setlength{\tabcolsep}{3pt}
\renewcommand{\arraystretch}{1.3}
\setlength{\baselineskip}{14pt}
\linespread{1.3}

\tableofcontents
\FloatBarrier

\section{\label{sec:level1} Parameters of the 16-qubit device}

\begin{figure}[t]
\centering
\includegraphics[width=1.0\textwidth]{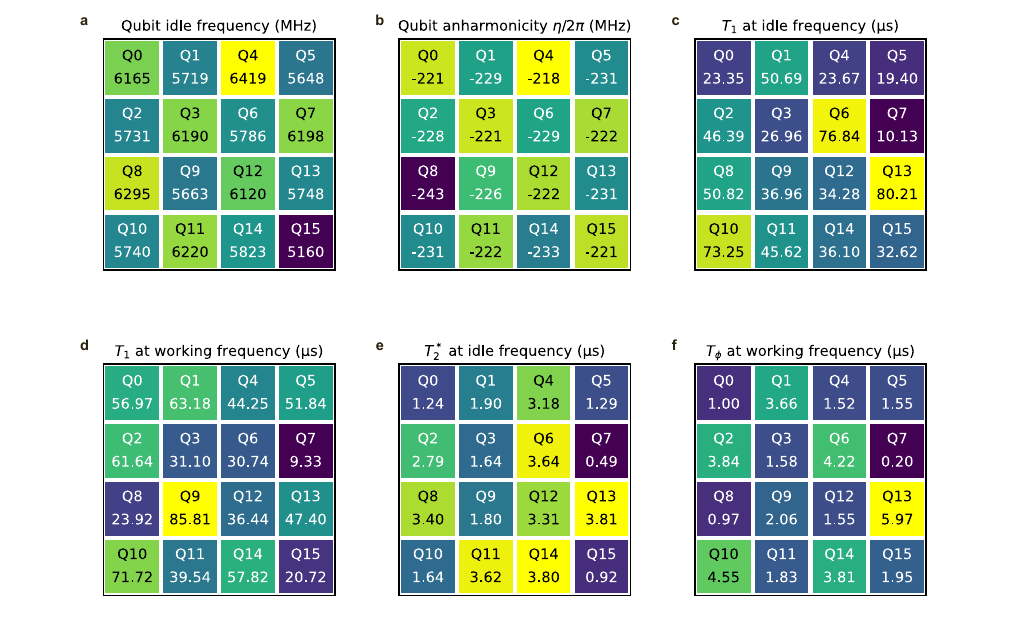}
\caption{\label{fig:qubit_parameter} \textbf{Qubit parameters.} \textbf{a},~Qubit idle frequency. \textbf{b},~Qubit anharmonicity $\eta$. \textbf{c},~$T_1$ at the idle frequency. \textbf{d},~$T_1$ at the working frequency. \textbf{e},~$T_2^*$ at the idle frequency. \textbf{f},~$T_{\phi}$ at the working frequency.} 
\end{figure}

\begin{figure}[t]
\centering
\includegraphics[width=1.0\textwidth]{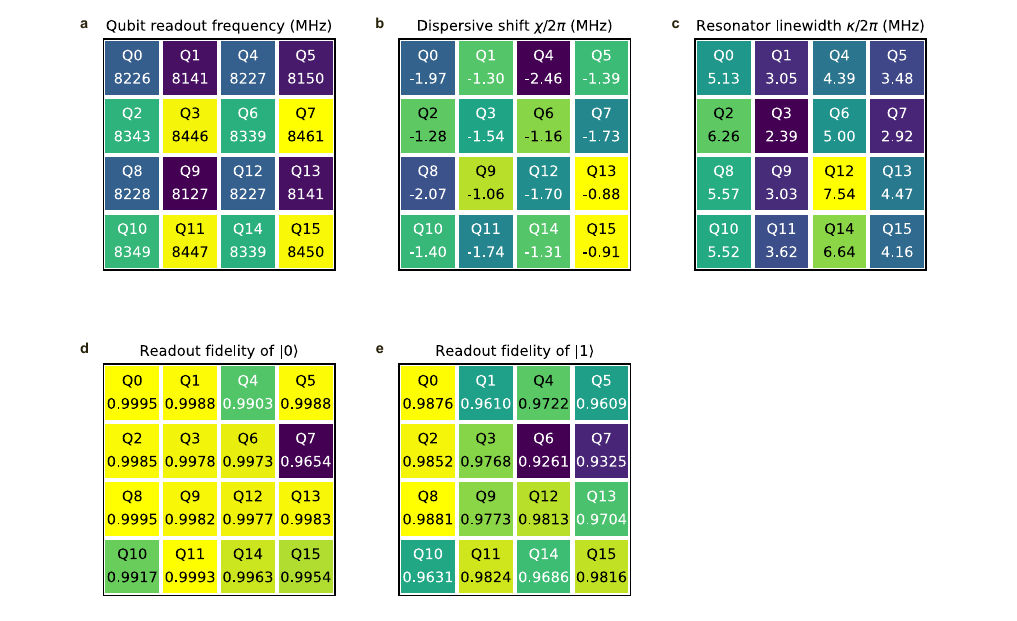}
\caption{\label{fig:read_parameter} \textbf{Readout parameters.} \textbf{a},~Qubit readout frequency. \textbf{b},~Dispersive shift $\chi$. \textbf{c},~Resonator linewidth $\kappa$. \textbf{d},~Readout fidelity of $\lvert 0\rangle$. \textbf{e},~Readout fidelity of $\lvert 1\rangle$.} 
\end{figure}

\begin{figure}[t]
\centering
\includegraphics[width=1.0\textwidth]{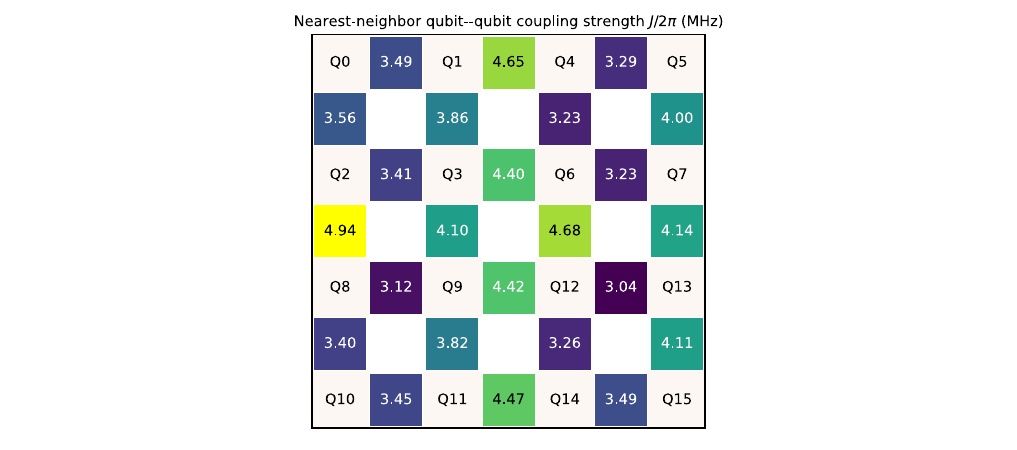}
\caption{\label{fig:coupling_parameter} \textbf{Nearest-neighbor qubit--qubit coupling strength $J$.} } 
\end{figure}

Our experiments are performed on a two-dimensional (2D) superconducting quantum processor comprising 16 frequency-tunable transmon qubits. The details of the processor and the measurement-system configuration can be found in a previous paper~\cite{yan2026characterizing}. Figure~\ref{fig:qubit_parameter} shows the parameters of the 16 qubits; Fig.~\ref{fig:read_parameter} shows the readout parameters of the 16 qubits; and Fig.~\ref{fig:coupling_parameter} shows the pairwise coupling strengths between the neighboring qubits. 

\section{\label{sec:level2} Calibration of the continuous measurement in the MZ interferometer}

\begin{figure}[t]
\centering
\includegraphics[width=1.0\textwidth]{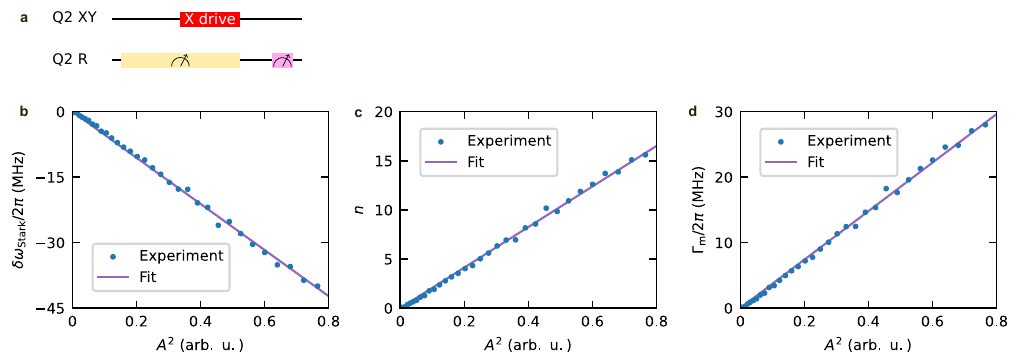}
\caption{\label{Stark_shift} \textbf{Stark-shift experiment of Q2 in the 12-qubit MZ interferometer.} \textbf{a},~Quantum circuit for the Stark-shift measurement. \textbf{b},~Stark shift $\delta \omega_\mathrm{Stark}$ of Q2. $A$ is the readout amplitude. \textbf{c},~Steady-state photon number $\bar{n}$ in the readout resonator during Q2 readout. \textbf{d},~Which-way measurement-induced dephasing rate $\Gamma_\mathrm{m}$ of Q2.} 
\end{figure}

In our experiment, we implement the which-way measurement by applying a pulse at the readout frequency to the readout resonator of Q2. The Hamiltonian describing the coupling between Q2 and its readout resonator is~\cite{blais2004cavity,gambetta2006qubit}: 
\begin{align} \label{Hqr}
	\hat H = \hbar\omega_\mathrm{r}\left(\hat a^{\dagger}\hat a+\frac{1}{2}\right)+\frac{\hbar\omega_\mathrm{q}}{2}\hat\sigma_z + \hbar g(\hat a^{\dagger}\hat \sigma_-+\hat a\hat \sigma_+),
\end{align}	
where $\omega_\mathrm{r}$ is the resonance frequency of the resonator, $\omega_\mathrm{q}$ is the qubit transition frequency, $\hat a^{\dagger}$($\hat a$) and $\hat \sigma_+$($\hat \sigma_-$) are the creation (annihilation) operators of the photon in the resonator and the qubit, and $g$ is the coupling strength between the qubit and the resonator.

When the qubit is largely detuned from the resonator, $|\Delta| = |\omega_\mathrm{q}-\omega_\mathrm{r}|\gg g$, the qubit and the resonator couple in the dispersive regime. Then we can perform a unitary transformation,
\begin{align} \label{U}
	U = \mathrm{exp}\left[\frac{g}{\Delta}(\hat a\hat \sigma_+-\hat a^{\dagger}\hat \sigma_-)\right],
\end{align}
to the Hamiltonian and expand to second order in $g$, yielding 
\begin{align} \label{UHU}
	UHU^{\dagger} = \hbar(\omega_\mathrm{r}+\chi\hat\sigma_z)\hat a^{\dagger}\hat a + \frac{\hbar}{2}(\omega_\mathrm{q}+\chi)\hat\sigma_z,
\end{align}
 where $\chi=g^2/\Delta$ is the dispersive shift. When the qubit's third level is taken into account~\cite{koch2007charge}, the dispersive shift becomes $g^{2}\eta/(\Delta(\Delta+\eta))$, where $\eta$ is the anharmonicity of the qubit. When the average photon number in the readout resonator is $\bar{n}$, the qubit Stark shift is given as $\delta\omega_\mathrm{{Stark}}=2\chi\bar{n}$ in the lowest-order dispersive approximation. When the measurement pulse is on resonance with the readout resonator, the measurement-induced dephasing rate can be written as~\cite{blais2021circuit}
\begin{align} \label{gamma_continuous}
	\Gamma_\mathrm{m} = \frac{2\kappa\chi^2\bar{n}}{\chi^2+(\kappa/2)^2}.
\end{align}

In the 12-qubit MZ interferometer experiment, we continuously apply the which-way measurement during the system evolution, and the average photon number in the readout cavity is kept constant in time. Therefore, we can extract the photon number and the measurement-induced dephasing rate by measuring the qubit Stark shift at different readout powers. As shown in Fig.~\ref{Stark_shift}a, we apply a continuous which-way measurement pulse to Q2 at the working frequency. After the readout-resonator photon number stabilizes, we apply a frequency-sweeping driving pulse to Q2 and measure Q2’s resonance frequency. Figure~\ref{Stark_shift}b shows that Q2’s Stark shift increases with the square of the readout amplitude $A$ (proportional to the readout power). We perform a linear fit constrained to pass through the origin. Figure~\ref{Stark_shift}c presents the steady-state photon number $\bar{n}$ in the readout resonator as a function of $A^2$. We vary the readout from the weak-measurement regime with $\bar{n} \ll 1$ to the strong-measurement regime with $\bar{n}$ around 16. Figure~\ref{Stark_shift}d shows the measurement-induced dephasing rate as a function of $A^2$. The dephasing rate $\Gamma_\mathrm{m}$ increases from much less than the coupling strength $J$ between neighboring qubits to values larger than $J$.

\section{\label{sec:level3} Calibration of the measurement for decoupled qubits in the MZ interferometer}

\begin{figure}[t]
\centering
\includegraphics[width=1.0\textwidth]{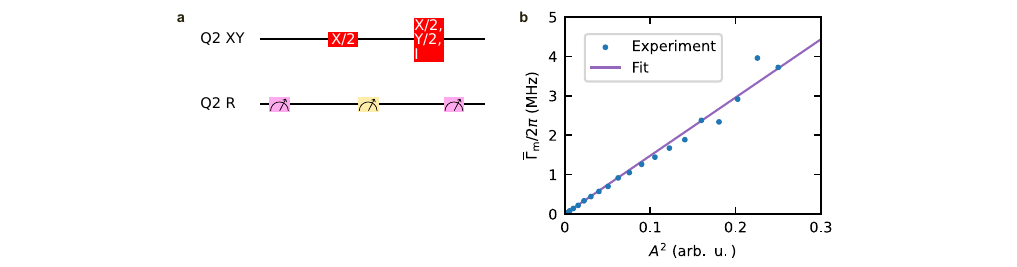}
\caption{\label{dephasing} \textbf{Measurement-induced dephasing rate of Q2 in the 4-qubit MZ interferometer.} 
\textbf{a},~Quantum circuit for measuring the average measurement-induced dephasing rate $\overline{\Gamma}_\mathrm{m}$.
\textbf{b},~$\overline{\Gamma}_\mathrm{m}$ of Q2 in the 4-qubit MZ interferometer as a function of the readout amplitude $A$.
} 
\end{figure}

\begin{figure}[t]
\centering
\includegraphics[width=1.0\textwidth]{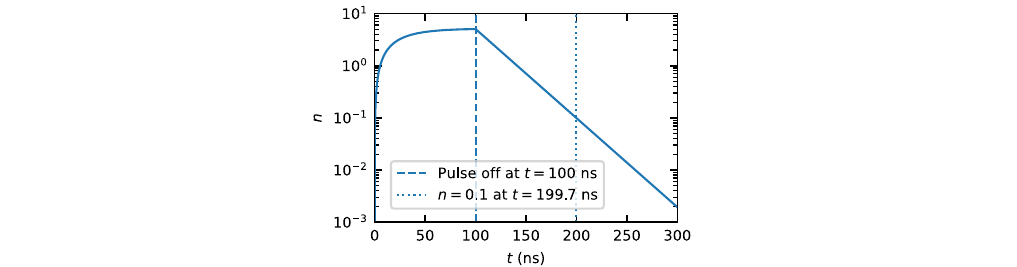}
\caption{\label{nphoton_4Q} \textbf{Calculation of the the time-dependent photon number $n(t)$ in Q2's readout resonator in the 4-qubit MZ interferometer.}
} 
\end{figure}

In the 4-qubit MZ interferometer, we decouple the qubits during the evolution and perform a which-way measurement on Q2. This measurement induces dephasing of Q2. As shown in Fig.~\ref{dephasing}a, after the initial measurement to post-select the initial state, we first apply an $X/2$ gate to prepare Q2 in state $\bigl(\lvert 0\rangle-i\lvert 1\rangle\bigr)/\sqrt2$.
Then we apply a 100-ns measurement pulse to Q2 and wait 200 ns for the readout-cavity photons to fully decay before performing state tomography on Q2. In the Markovian master equation, the off-diagonal element $\rho_\mathrm{01}$ of Q2’s density matrix decays exponentially with time constant $T_2$. We denote the off-diagonal element without the which-way measurement by $\rho_\mathrm{01}^{\mathrm{ref}}$ and with the which-way measurement by $\rho_\mathrm{01}^{\mathrm{m}}$. The average measurement-induced dephasing rate of Q2 is then given by
\begin{align} \label{gamma_single_time}
	\overline{\Gamma}_\mathrm{m} = -\dfrac{1}{t_\mathrm{m}}\ln\!\left(\dfrac{\lvert\rho_\mathrm{01}^{\mathrm{m}}\rvert}{\lvert\rho_\mathrm{01}^{\mathrm{ref}}\rvert}\right),
\end{align}
where $t_{m}$ is the measurement time. As shown in Fig.~\ref{dephasing}b, $\overline{\Gamma}_\mathrm{m}/2\pi$ increases linearly with $A^2$. 

In our readout model, the cavity photon number $n(t)$ follows a driven-dissipative evolution: during the readout pulse ($t=0$--100~ns) the intra-cavity population approaches the steady-state value \(n_{\mathrm{ss}}\) according to
\begin{align} \label{nphoton_drive}
n(t)=n_{\mathrm{ss}}\left[1-\exp(-\kappa t)\right],
\end{align}
and after the pulse is turned off the population decays exponentially as
\begin{align} \label{nphoton_off}
n(t)=n(t_{\mathrm{off}})\,\exp\!\left[-\kappa (t-t_{\mathrm{off}})\right].
\end{align}
As the parameters shown in Fig.~\ref{fig:read_parameter}, we set \(\kappa/2\pi=6.26\ \mathrm{MHz}\).
In the 4-qubit MZ interferometer experiment, the maximum readout strength we use is \(A^{2}=0.25\). From Fig.~\ref{Stark_shift}b we observe that \(A^{2}=0.25\) corresponds to a resonator photon number of \(5.16\). Figure~\ref{nphoton_4Q} shows the simulated cavity photon-number evolution during the which-way measurement. According to Eq.~\eqref{nphoton_drive}, during the applied 100-ns readout pulse, the intra-cavity photon number rises from \(0\) to \(5.06\). According to Eq.~\eqref{nphoton_off}, after the readout pulse is turned off, the photon number decays exponentially with rate \(\kappa\), and it falls below \(0.1\) after $t=199.7$~ns. We therefore take the measurement time \(t_\mathrm{m}\) in the sequence to be \(200\ \mathrm{ns}\).

\section{\label{sec:level4} Numerical simulation}

The time evolution of our multi-qubit system is described by a Lindblad master equation. The density matrix $\rho$ evolves as
\begin{equation}\label{eq:lindblad}
\dot{\rho}(t) \;=\; -\frac{i}{\hbar}\big[\hat H(t),\,\rho(t)\big]
\;+\; \sum_{n}\frac{1}{2}\Big[2\hat C_{n}\,\rho(t)\,\hat C_{n}^{\dagger}
-\rho(t)\,\hat C_{n}^{\dagger}\hat C_{n}-\hat C_{n}^{\dagger}\hat C_{n}\,\rho(t)\Big],
\end{equation}
where $\hat H$ is the system Hamiltonian, and $\hat C_{n} \;=\; \sqrt{\gamma_{n}}\,\hat A_{n}$ are the collapse operators; $\hat A_{n}$ are the operators through which the environment couples to the system, and $\gamma_{n}$ are the corresponding rates.
The Hamiltonian of our system can be described by a Bose-Hubbard model:
\begin{align} \label{H}
	\hat H = \sum _{i}\left[\hbar\omega_i\hat a_i^{\dagger}\hat a_i+\frac{\hbar U_i}{2}\hat n_i(\hat n_i-1)\right] + \sum _{\langle i,j\rangle}\hbar J_{ij}(\hat a_i^{\dagger}\hat a_j+\hat a_i\hat a_j^{\dagger}).
\end{align}	

We model dissipation in QuTiP using collapse operators passed to the \texttt{qutip.mesolve} function~\cite{johansson2012qutip,LAMBERT20261}. We denote the collapse operator that describes the decay of qubit $i$ by $\hat c_{\text{ops1},i}$ and the collapse operator that describes the dephasing of qubit $i$ by $\hat c_{\text{ops2},i}$. Concretely, we use
\begin{align}
\hat c_{\text{ops1},i} &= \sqrt{\Gamma_{1,i}}\,\hat \sigma_{i}^{-}, \qquad \Gamma_{1,i} = 1/T_{1,i}, \\
\hat c_{\text{ops2},i} &= \sqrt{\Gamma_{\phi,i}/2}\,\hat \sigma_{i}^{z}, \qquad \Gamma_{\phi,i} = 1/T_{2,i}+\Gamma_\mathrm{m},
\end{align}
where $T_{1,i}$ and $T_{2,i}$ are set to the values measured at the corresponding operating point shown in Fig.~\ref{fig:qubit_parameter}.

\paragraph{\textbf{12-qubit continuous-measurement case.}}
For the continuous-measurement case of the 12-qubit MZ interferometer, we initialize the system in the state $\lvert1000\cdots\rangle$ (i.e., only Q0 is excited). We set all qubit frequencies to the working frequency, and set $T_{1}$ and $T_{2}$ to the values measured at the working point. Then we evolve the system according to Eq.~\eqref{eq:lindblad}. For each measurement strength, we take the measurement-induced dephasing rate $\Gamma_\mathrm{m}$ of Q2 obtained in Fig.~\ref{Stark_shift}b, update $\Gamma_{\phi,2}$ accordingly, and include the updated values in the collapse operators $\hat c_{\text{ops2},2}$ when solving the master equation. From the density matrix of the final state we compute the populations, the two-qubit concurrence, the single-excitation subspace purity and von Neumann entropy, and other quantities.

\paragraph{\textbf{4-qubit MZ interferometer simulation (which-way measurement for decoupled qubits).}}
For the simulation of the 4-qubit MZ interferometer, we set the initial state to $\lvert1000\rangle$. We divide the evolution into four successive stages and solve the master equation separately for each stage. We use the final 4-qubit state of each stage as the initial state of the next. For each qubit we set $T_{1}$ and $T_{2}$ to the values measured at the corresponding frequency. The four stages are:

\begin{itemize}
  \item \textbf{Stage 1 — photon generation and beamsplitting.} The frequencies of the 4 qubits are set to [5.820, 5.820, 5.820, 6.190] GHz. We align Q0, Q1, Q2 for joint evolution for $57\ \mathrm{ns}$.
  \item \textbf{Stage 2 — which-way measurement.} The frequencies of the 4 qubits are set to [6.165, 5.830, 5.820, 6.190] GHz. We include the effect of the which-way measurement by updating $\Gamma_{\phi,2}$ according to the average measurement-induced dephasing rate $\overline{\Gamma}_\mathrm{m}/2\pi$ obtained in Fig.~\ref{dephasing}b and include the corresponding $\hat c_{\text{ops2},2}$ term. The evolution time is $200\ \mathrm{ns}$.
  \item \textbf{Stage 3 — phase accumulation on path 1.} Frequencies of the 4 qubits are set to [6.165, 5.830 + $\delta/2\pi$, 5.820, 6.190] GHz. We scan $\delta$ to change Q1's frequency so that the photon accumulates a phase $\phi = - t\delta$ at Q1. We evolve for $100\ \mathrm{ns}$.
  \item \textbf{Stage 4 — interference.} Frequencies of the 4 qubits are set to [6.165, 5.820, 5.820, 5.820] GHz. We align Q1, Q2, Q3 for joint evolution for $49\ \mathrm{ns}$, after which the photon interferes at Q3.
\end{itemize}

From the final density matrix we extract the probability $P_\mathrm{I}$ of Q3 being in the $\lvert1\rangle$ state as a function of $\delta$, yielding the interference fringes.

\paragraph{\textbf{Simulation of Q1--Q2 tomography.}}
To simulate the tomography results for the Q1--Q2 two-qubit subsystem (which exclude the phase-accumulation and final interference steps), we perform only Stage~1 and Stage~2. We replace Stage~3 with the evolution with the 4 qubits at frequency [6.165, 5.830, 5.820, 6.190] GHz for $100\ \mathrm{ns}$. From the resulting final state we compute the two-qubit concurrence, the single-excitation subspace purity and von Neumann entropy, and other quantities.

\section{\label{sec:level5} Qubit-frequency alignment}

Our experiment requires precise control of the frequencies of all qubits in the system: we need to align qubit frequencies at the working point, eliminate DC and $Z$-pulse crosstalk, and stabilize the qubit frequencies for long periods. In our previous work~\cite{yan2026characterizing} all qubits were aligned to the working point using DC bias. For systems with many qubits, applying DC bias to tune every qubit to the working point may require large DC currents for some qubits and can cause heating due to the added infrared filter at the mixing-chamber stage. To make our approach scalable, in this work we use $Z$ pulses to bias qubits to the working and idle points. We realize precise frequency control and alignment of the qubits via the following procedure:

\begin{enumerate}
  \item Measure the DC-bias-dependent spectrum of each qubit and choose a modest DC bias that places each qubit at a suitable frequency. This choice ensures that the system does not overheat while leaving sufficient range to further bias each qubit to the working and idle points using $Z$ pulses.
  \item Precisely measure the $Z$-pulse-dependent spectroscopy of each qubit at the chosen DC-bias point with using $Z$ pulses to bias each qubit to its idle frequency and correct $Z$-pulse distortion. Then perform Ramsey measurements to finely tune the $Z$-pulse amplitude $A_{\mathrm{idle}}$ so that each qubit sits at its idle point. For qubits that are not used in the experiment, we bias them to $5.620\ \mathrm{GHz}$ or $6.020\ \mathrm{GHz}$ by $Z$ pulses.
  \item Apply $Z$-pulse biases to all qubits simultaneously and, for each qubit in turn, perform Ramsey measurements to finely adjust the $Z$-pulse amplitude $A_{\mathrm{idle}}'$ to remove $Z$-pulse crosstalk.
  \item Apply $Z$ pulses to all qubits simultaneously to bias them to the working point for evolution. Compare the experimental evolution with numerical simulation and optimize the $Z$-pulse amplitudes $A_{\mathrm{working}}$ using the Nelder--Mead algorithm to minimize the effect of $Z$-pulse crosstalk and achieve precise frequency alignment for all qubits.
  \item After completing calibration steps 1--4, we begin the experiments. During the experiments, magnetic-field changes or DC drift can shift qubit frequencies. We perform a qubit-frequency calibration approximately every $2\ \mathrm{hours}$. For each qubit we apply a $Z$ pulse to bias it to its idle point and perform a Ramsey experiment. If a frequency shift is observed, we correct the qubit idle frequency by finely adjusting the DC bias while keeping $A_{\mathrm{idle}}$, $A_{\mathrm{idle}}'$, and $A_{\mathrm{working}}$ unchanged. This procedure enables rapid frequency correction without redoing the full alignment.
\end{enumerate}

\paragraph{Details for the frequency-alignment procedure in step 4.}
\begin{enumerate}
  \item \textbf{4-qubit MZ interferometer:} 
  For the 4-qubit MZ interferometer we perform separate alignments for the Q0, Q1, Q2 subsystem and the Q1, Q2, Q3 subsystem.

  To align Q0, Q1, Q2, we initially excite Q0 to $\lvert1\rangle$ and evolve the system until the Q1 and Q2 excitation probabilities $P_\mathrm{Q1}(1)$ and $P_\mathrm{Q2}(1)$ reach their maxima. We define the cost function as the sum of the differences between the experimental and simulated populations at that time,
  \begin{align} \label{costf_4Q1}
    C = \big|P_\mathrm{Q1}^{\mathrm{exp}}(1)-P_\mathrm{Q1}^{\mathrm{sim}}(1)\big| + \big|P_\mathrm{Q2}^{\mathrm{exp}}(1)-P_\mathrm{Q2}^{\mathrm{sim}}(1)\big|,
  \end{align}
  and optimize the $Z$-pulse amplitudes to minimize $C$, thereby achieving alignment.

  To align Q1, Q2, Q3, we initially excite Q1 and Q2 to $\lvert1\rangle$ and evolve until $P_\mathrm{Q3}(1)$ reaches its maximum. We use the difference between the experimental and simulated $P_\mathrm{Q3}(1)$ at that time as the cost function,
  \begin{align} \label{costf_4Q2}
    C = \big|P_\mathrm{Q3}^{\mathrm{exp}}(1)-P_\mathrm{Q3}^{\mathrm{sim}}(1)\big|,
  \end{align}
  and optimize the $Z$-pulse amplitudes to minimize $C$.

  \item \textbf{12-qubit MZ interferometer:} 
  For the 12-qubit MZ interferometer we initially excite Q0 to $\lvert1\rangle$ and evolve the system until the excitation probabilities of Q4, Q8, and Q15 reach their maxima, respectively. We define the cost function as the sum of the differences between the experimental and simulated populations of these qubits,
  \begin{align} \label{costf_12Q}
    C = \big|P_\mathrm{Q4}^{\mathrm{exp}}(1)-P_\mathrm{Q4}^{\mathrm{sim}}(1)\big| + \big|P_\mathrm{Q8}^{\mathrm{exp}}(1)-P_\mathrm{Q8}^{\mathrm{sim}}(1)\big| + \big|P_\mathrm{Q15}^{\mathrm{exp}}(1)-P_\mathrm{Q15}^{\mathrm{sim}}(1)\big|,
  \end{align}
  and optimize the $Z$-pulse amplitudes to minimize $C$, thereby achieving qubits alignment.
\end{enumerate}

\section{Derivation of complementarity relations}

We obtain the effective single-excitation description by projecting the reconstructed two-qubit density matrix onto the subspace spanned by the basis states \(\lvert 10\rangle\) and \(\lvert 01\rangle\).  Denoting the full Q1–Q2 density matrix by $\rho_\mathrm{Q1Q2}$~(ordered as \(\{\lvert 00\rangle,\lvert 01\rangle,\lvert 10\rangle,\lvert 11\rangle\}\)) and the single-excitation projector by
\begin{align} \label{projector}
\hat P \;=\; \lvert 10\rangle\langle 10\rvert + \lvert 01\rangle\langle 01\rvert,
\end{align}
we define the single-excitation weight
\begin{align} \label{weight}
s \;=\; \mathrm{Tr}\bigl(\hat P\,\rho_\mathrm{Q1Q2}\bigr)
\end{align}
and the single-excitation density matrix
\begin{align} \label{rhos}
\rho_\mathrm{s} \;=\; \frac{\hat P\,\rho_\mathrm{Q1Q2}\,\hat P}{s}
\qquad (s>0).
\end{align}
Writing the matrix elements of \(\rho_\mathrm{s}\) in the ordered basis \(\{\lvert 10\rangle,\lvert 01\rangle\}\) as
\begin{align} \label{rhos2}
\rho_\mathrm{s} \;=\; \begin{pmatrix} p_1' & c' \\[4pt] {c'}^{*} & p_2' \end{pmatrix}
\qquad (p_1' + p_2' = 1),
\end{align}
we use the notation $p_1' = p_1/s$, $p_2' = p_2/s$ and $c' = c/s$ where $p_{1}=\rho_\mathrm{Q1Q2,10,10}$, $p_{2}=\rho_\mathrm{Q1Q2,01,01}$ and $c=\rho_\mathrm{Q1Q2,10,01}$ denote elements of the full $4\times4$ matrix $\rho_\mathrm{Q1Q2}$. The interference pattern observed at Q3 is determined by the off-diagonal coherence \(c'\) and the populations $p_1'$ and $p_2'$. We can obtain the expression for the visibility:
\begin{align} \label{visibility}
V \;=\; 2\lvert c'\rvert.
\end{align}

For a general $2\times2$ density matrix of this form, let $\delta p'\equiv p_1' - p_2'$.  The eigenvalues of $\rho_\mathrm{s}$ read
\begin{align} \label{lambda}
\lambda_{\pm} \;=\; \frac{1}{2}\Bigl(1 \pm R\Bigr), 
\end{align}
where $R \equiv \sqrt{\delta p'^2+V^2}$ $(0\le R \le 1)$. The purity $P_\mathrm{s}$ in the single-excitation subspace is given as
\begin{align}\label{purity}
P_\mathrm{s} = \operatorname{Tr}(\rho_\mathrm{s}^2) \;=\; \lambda_+^2 + \lambda_-^2
= \frac{1+R^2}{2}. 
\end{align}
Thus, we can obtain the complementarity relation
\begin{align}
2(1-P_\mathrm{s}) + V^{2}
&= 1 - R^{2} + V^{2}
= 1 - (\delta p'^{2}+V^{2}) + V^{2} \\
&= 1 - \delta p'^{2} \;\le\; 1,
\end{align}
where equality holds if and only if \(\delta p'=0\) (i.e.\ the two paths are population-balanced). Note that $(1-P_\mathrm{s})$ equals the linear entropy and as such gives an alternative measure of the mixedness of the quantum state.

The von Neumann entropy in the single-excitation subspace is
\begin{align} \label{Ss}
S_\mathrm{s} \;=\; -\sum_{i=\pm}\lambda_{i}\,\log_{2}\lambda_{i}
\;=\; -\biggl(\frac{1+R}{2}\biggr)\log_{2}\biggl(\frac{1+R}{2}\biggr) \;-\; \biggl(\frac{1-R}{2}\biggr)\log_{2}\biggl(\frac{1-R}{2}\biggr).
\end{align}
The expression above shows that, for fixed visibility $V$, the entropy $S_\mathrm{s}$ is maximized when $\delta p'=0$ (i.e., $p_1'=p_2'$).  Denoting the symmetric (population-balanced) entropy by
\begin{align} \label{Ssym}
S_\mathrm{s}^\mathrm{sym}(V) \;=\; -\biggl(\frac{1+V}{2}\biggr)\log_{2}\biggl(\frac{1+V}{2}\biggr) \;-\; \biggl(\frac{1-V}{2}\biggr)\log_{2}\biggl(\frac{1-V}{2}\biggr),
\end{align}
we obtain the general inequality 
\begin{align} \label{ineq}
0 \;\le\; S_\mathrm{s} \;\le\; S_\mathrm{s}^\mathrm{sym}(V),
\end{align}
where the upper bound is achieved if and only if the two paths are population-balanced ($p_1'=p_2'$), and the lower bound $S_\mathrm{s}=0$ is achieved if and only if the state is pure ($R=1$).  

Physically, projecting onto the single-excitation subspace restricts the system to a single photon, and the resulting \(\rho_\mathrm{s}\) can be mapped to the density matrix of a single photon in a two-path MZ interferometer. The which-way measurements reduce the value of \(\lvert c'\rvert\) and hence \(V\); while $(1-P_\mathrm{s})$ and $S_\mathrm{s}$ quantify the loss of coherence (mixedness) that accompanies the reduction of visibility.

\section{ Analysis of the measurement process of which-way measurements}

In circuit quantum electrodynamics the dispersive readout maps the qubit state
onto two conditional resonator pointer states, $|\alpha_g(t)\rangle$ and
$|\alpha_e(t)\rangle$~\cite{gambetta2006qubit}. Under the usual dispersive Hamiltonian
\begin{equation}
\frac{\hat H}{\hbar} = \omega_\mathrm{r} \hat a^\dagger \hat a + \frac{\omega_q}{2}\hat \sigma_z + \chi\, \hat a^\dagger \hat a\,\hat \sigma_z,
\end{equation}
and for a driven, damped resonator with decay rate $\kappa$ the pointer amplitudes
satisfy~\cite{gambetta2008quantum}
\begin{equation}
\dot{\alpha}_{s}(t) = -\left[\frac{\kappa}{2}+i\big(\Delta + s\chi\big)\right]\alpha_s(t) - i\varepsilon_\mathrm{d}(t),
\end{equation}
where $s\in\{+1,-1\}$ for $\{|e\rangle,|g\rangle\}$, $\Delta=\omega_\mathrm{r}-\omega_{\mathrm{drive}}$ is the difference between the resonator frequency $\omega_\mathrm{r}$ and the drive frequency $\omega_{\mathrm{drive}}$, and $\varepsilon_\mathrm{d}(t)$ is the amplitude of the external drive. The distinguishability of the
two qubit-conditioned fields is governed by $\delta\alpha = \alpha_e(t)-\alpha_g(t)$.
When the qubits are decoupled during the readout interval $[0,T]$, the effective qubit Hamiltonian commutes with the measured observable, and the measurement reduces to a single-shot discrimination of Gaussian-distributed detector outcomes. This is the situation encountered in the 4-qubit experiment in the main text. Denoting the integrated detector result
by $s$, the conditional probabilities take the Gaussian forms,
\begin{eqnarray}
p(s\mid g) &=& \frac{1}{\sqrt{2\pi\sigma^2}} \exp\!\left[-\frac{(s-s_g)^2}{2\sigma^2}\right], \\
p(s\mid e) &=& \frac{1}{\sqrt{2\pi\sigma^2}} \exp\!\left[-\frac{(s-s_e)^2}{2\sigma^2}\right],
\end{eqnarray}
where $s_{g(e)}$ are the mean integrated signals when the qubit is in $|g\rangle(|e\rangle)$ and $\sigma^2 \propto 1/T$ is the detector variance after the integration time $T$. The resulting positive operator-valued measure (POVM) element associated with the measurement outcome $s$ is diagonal in the
energy basis and reads
\begin{equation}
E(s)= p(s\mid g)\,|g\rangle\langle g| + p(s\mid e)\,|e\rangle\langle e|,
\label{eq:POVM_frozen}
\end{equation}
which satisfies $E(s)\ge 0$ and $\int ds\, E(s) = I$.
This diagonal form reflects the QND character of dispersive readout in the
decoupled-qubit limit and supports a straightforward Bayesian update of the
population probabilities. In the limit of large measurement effect $\Gamma_m T$, where $\Gamma_m$ denotes the measurement rate, the signal-to-noise ratio becomes large, and the above POVM coarse-grains to projections onto
$|g\rangle$ and $|e\rangle$. For weak and intermediate measurement effects, the distributions $p(s \mid g)$ and $p(s \mid e)$ overlap and the POVM cannot discriminate the two states with certainty, which allows us in the main text to explore the complementarity between which-way information and visibility in a controlled way.

In contrast, when the qubits are coupled during the readout interval, the effective qubit Hamiltonian does not commute with the measured observable, and the measurement outcome is a full stochastic record $\{I(t)\}_{0\le t\le T}$ rather than a single scalar $s=T^{-1}\int_0^T I(t)\,dt$. This is the situation encountered in the 12-qubit experiment in the main text. The conditioned density matrix $\rho_I(t)$ then obeys a stochastic master equation of the form~\cite{wiseman2009quantum, jacobs2014quantum}
\begin{equation}
d\rho_I = -i[\hat H_\mathrm{q},\rho_I]\,dt + \Gamma_\mathrm{m}\,\mathcal{\hat D}[\hat A]\rho_I\,dt + \sqrt{\Gamma_\mathrm{m}}\,\mathcal{\hat H}[\hat A]\rho_I\,dW(t),
\label{eq:SME}
\end{equation}
with $\mathcal{\hat D}[\hat A]\rho = \hat A\rho \hat A^\dagger -\tfrac{1}{2}\{\hat A^\dagger \hat A,\rho\}$, $\mathcal{\hat H}[\hat A]\rho = \hat A\rho+\rho \hat A^\dagger - \operatorname{Tr}\big[(\hat A\rho+\rho \hat A^\dagger)\big]\,\rho$, $\hat A$ the measured system operator (proportional to $\sigma_z$ in the
dispersive readout), and $\Gamma_m$ the measurement rate. The Wiener increment $dW$ is related to the measured signal through
\begin{align}
I(t) dt = \langle \hat{A} \rangle_I dt + dW .
\label{eq:measurement_record} 
\end{align}
The solution of the stochastic master equation (\ref{eq:SME}) describes the measurement backaction for a given initial state $\rho_0$ and associated measurement record (\ref{eq:measurement_record}), while the statistics of the Wiener increment with $\mathbb{E}[dW]=0$ and $\mathbb{E}[dW^2]=dt$ determines the probability to obtain the signal $I(t)$. In the 12-qubit experiment of the main text, the noncommutativity of $\hat H_\mathrm{q}$ and $\hat A$ gives rise to the quantum Zeno effect, where (partial) measurement-induced reflection at the monitored qubit redistributes the wave components between the two MZ interferometer arms, resulting in a modified complementarity response.

\bibliographystyle{rev4-2mod}
\bibliography{main}